\documentclass[aps,twocolumn,showpacs,preprintnumbers,prb,amsmath,amssymb,amsfonts,superscriptaddress,floatfix]{revtex4-1}

\usepackage{graphicx}
\usepackage{bm}

\begin{document}

\title{Tunability of electronic and optical properties of the
Ba-Zr-S system via dimensional reduction}

\author{Yuwei Li}

\author{David J. Singh}
\email{singhdj@missouri.edu}

\affiliation{Department of Physics and Astronomy, University of Missouri,
Columbia, MO 65211-7010, USA}

\date{\today}

\begin{abstract}
Transition metal sulfide perovskites offer lower band gaps and
greater tunability than oxides,
along with other desirable properties for applications. Here we explore
dimensional reduction as a tuning strategy using the Ruddlesden-Popper
phases in the Ba-Zr-S system as a model. The three dimensional
perovskite BaZrS$_3$ is a direct gap semiconductor, with a band gap
of 1.5 eV suitable for solar photovoltaic application. However, the
three known members of the Ruddlesden-Popper series, are all indirect
gap materials, and additionally have lower fundamental band gaps.
This is accompanied in the case of Ba$_2$ZrS$_4$
by a band structure that is more favorable
for carrier transport for oriented samples.
The layered Ruddlesden-Popper compounds show significantly anisotropic
optical properties, as may be expected.
The optical spectra show tails at low energy,
which may complicate experimental characterization of these materials.
\end{abstract}

\maketitle

\section{Introduction}

The sulfide perovskite BaZrS$_3$ has a moderate band gap of less than 2 eV
and exhibits ambipolar doping. \cite{perera,niu,sun,meng}
This is in contrast to the corresponding oxide, BaZrO$_3$ (band gap
5.3 eV). \cite{robertson}
Furthermore, there is evidence that the band gap of BaZrS$_3$
might be considerably tuned by substitutions on the normally inactive
$A$-site of the $ABX_3$ perovskite structure,
\cite{niu,niu2} also different from most semiconducting perovskite oxides.
Tunability of electronic and optical properties is an important consideration
for electronic and optoelectronic applications. This is exemplified
by the wide tunability in III-V and II-VI zinc blende structure
semiconductor alloys.
These alloys thus have a multitude of
ensuing applications and are the basis of key technologies.
In fact, it is known that the properties of sulfide perovskites can be
modified by alloying, for example with Ti, but the lack of stability of the
Ba(Zr,Ti)S$_3$ alloy is a challenge. \cite{meng}
In addition to transition metal alloying,
substitutions for Ba on the perovskite $A$-site,
and chalcogen alloys are also possible and may be effective in tuning
the electronic and optical properties. However, alternatives are
desirable.

Dimensional reduction is
another highly effective general method for modifying electronic structures
of semiconductors.
This can be done by artificial methods, such as through quantum wells
and superlattices, which are important for semiconductor technologies.
\cite{kroemer}
Additionally, the electronic structure changes associated with 
dimensional reduction may be important for a wide variety of
applications, including for example, thermoelectrics, where
they can decouple the thermopower and conductivity,
\cite{hicks,zhang2,xing,terasaki}
optoelectronic materials, such as transparent conductors,
\cite{li-basno3}
and a wide variety of optical, electronic and other applications based
on 2D materials. \cite{wang,butler}

Dimensional reduction in the electronic structure of materials can be
achieved in various ways. These include artificial structures, such
as superlattices, exfoliation or thin film growth to produce
2D materials, interfacial electron gasses,
\cite{ohtomo}
bonding structures that lower the effective dimensionality,
\cite{parker}
and the production of homologous series of layered compounds,
such as the Ruddlesden-Popper (RP) series for perovskites.
\cite{ruddlesden}
This series consists of perovskite structure blocks separated by 
extra rock salt layers along a [001] direction. The first
element of this series is the K$_2$NiF$_4$ structure, which is the
prototype structure of the high temperature superconductors.

In the case of BaZrS$_3$ three members of the
RP series, Ba$_{n+1}$Zr$_n$S$_{3n+1}$, where $n$
is the number of layers in the perovskite block,
are known experimentally.
\cite{lelieveld,saeki,chen,chen2}
These are the $n$=1, 2 and 3 members, Ba$_2$ZrS$_4$,
Ba$_3$Zr$_2$S$_7$ and Ba$_4$Zr$_3$S$_{10}$, in addition to the
$n$=$\infty$ perovskite, BaZrS$_3$.
The 3D perovskite, BaZrS$_3$ and the $n$=3 compound, Ba$_4$Zr$_3$S$_{10}$
show significant distortions from the ideal
structure due to rotation of the ZrS$_6$ octahedra,
consistent with the fact that the perovskite tolerance factor,
$t$=0.95 (based on Shannon crystal radii)\cite{shannon}
is smaller than unity for these compounds,
while the $n$=1 and $n$=2 compounds have been reported to
occur in an ideal undistorted body centered tetragonal structure.

The purpose of this paper is to report a consistent set of
first principles electronic structures and optical properties
for this series of compounds.
We find that dimensional reduction lowers the band gap in contrast
to most oxide perovskites, and that additionally it changes
the band gap character from direct in the 3D perovskite to
indirect in the layered RP compounds. We also find that
dimensional reduction can substantially enhance carrier transport
for both electrons and holes in suitably oriented samples.

\section{Structure and Methods}

\begin{figure}
\includegraphics[width=\columnwidth,angle=0]{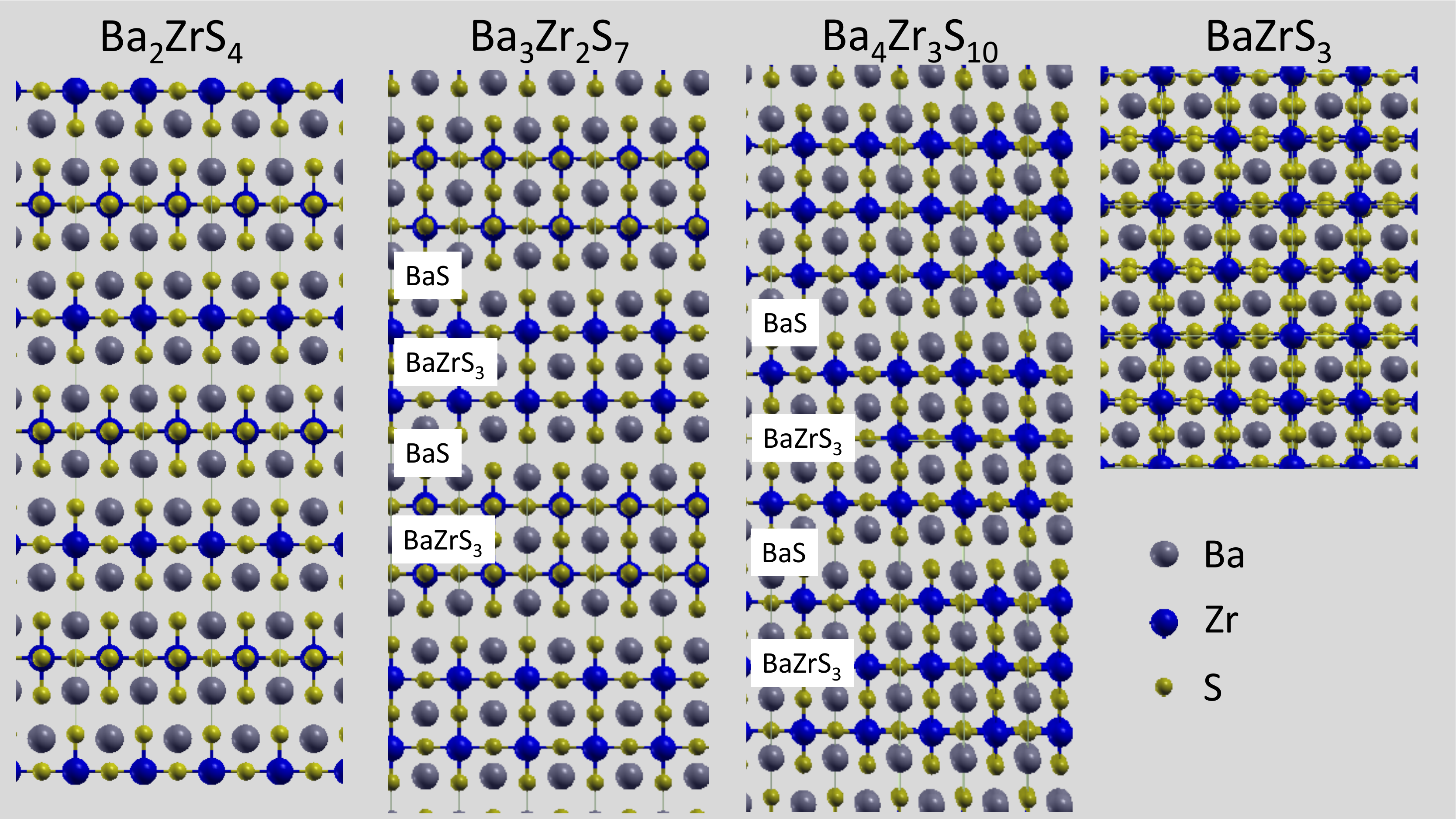}
\caption{Structures of the compounds. These include the first principles
relaxed atomic coordinates as described in the text. The rock salt,
BaS, and perovskite parts of the cell are labeled for the
Ba$_3$Zr$_2$S$_7$ and Ba$_4$Zr$_3$S$_{10}$ compounds.}
\label{struct}
\end{figure}

Our calculations were done within density functional
theory (DFT). The electronic structures and optical properties were
calculated using the general potential linearized augmented planewave (LAPW)
method, \cite{singh-book}
as implemented in the WIEN2k code. \cite{wien2k}
The total energy calculations and
relaxation of the atomic coordinates were done using the PBE
generalized gradient approximation. \cite{pbe}
For these relaxations,
relativity was treated at a scalar relativistic level, and relaxation was
continued until all force components
were below 1 mRy/Bohr ($\sim$0.025 eV/\AA).
LAPW sphere radii of 2.5 bohr, for Ba,
2.3 Bohr for S and 2.3 Bohr, for Zr were used, except for Ba$_2$ZrS$_4$,
for which the S radius was reduced to 2.2 Bohr to accommodate the shorter
bond length in this compound.
The basis sets were defined by
a planewave sector basis cutoff determined by $R_{min}K_{max}$=9,
where $R_{min}$ is the smallest LAPW sphere radius.
The standard LAPW augmentation was employed.
Local orbitals were added to the basis to treat semicore states.
We used the experimental lattice parameters from literature, and
relaxed all internal atomic coordinates by total energy minimization.
Specifically, we used
space group $Pnma$, $a$=7.0599 \AA, $b$=9.9813 \AA, $c$=7.0251 \AA, 
for BaZrS$_3$, \cite{lelieveld}
space group $I4/mmm$, $a$=4.7852 \AA, $c$=15.9641 \AA,
for Ba$_2$ZrS$_4$, \cite{saeki}
space group $I4/mmm$, $a$=4.9983 \AA, $c$=25.502 \AA,
for Ba$_3$Zr$_2$S$_7$, \cite{chen}
and
space group $Fmmm$, $a$=7.0314 \AA, $b$=7.0552 \AA, $c$=35.544 \AA, 
for Ba$_4$Zr$_3$S$_{10}$. \cite{chen2}
In the case of Ba$_3$Zr$_2$S$_7$, we also performed calculations for the
orthorhombic structure reported by Saeki, \cite{saeki}
but find that it is not energetically favored.
The crystal structures are depicted in Fig. \ref{struct}.
The relaxed atomic coordinates are given in Tab. \ref{struct-tab}.

Spin-orbit was included for the electronic and optical properties.
The band gap is important for these, and accordingly these
calculations were done using the modified Becke-Johnson potential of
Tran and Blaha. \cite{mbj}
This functional gives band gaps in good accord with experiment for a wide
variety of simple semiconductors and insulators.
\cite{mbj,koller-mbj,singh-mbj}
Calculation of the transport function for conductivity was done using
the BoltzTraP code. \cite{boltztrap}
Optical properties were calculated based on electric dipole transitions
in the independent particle approximation as implemented in the WIEN2k code.

\section{Results and Discussion}

\begin{figure}
\includegraphics[width=0.95\columnwidth,angle=0]{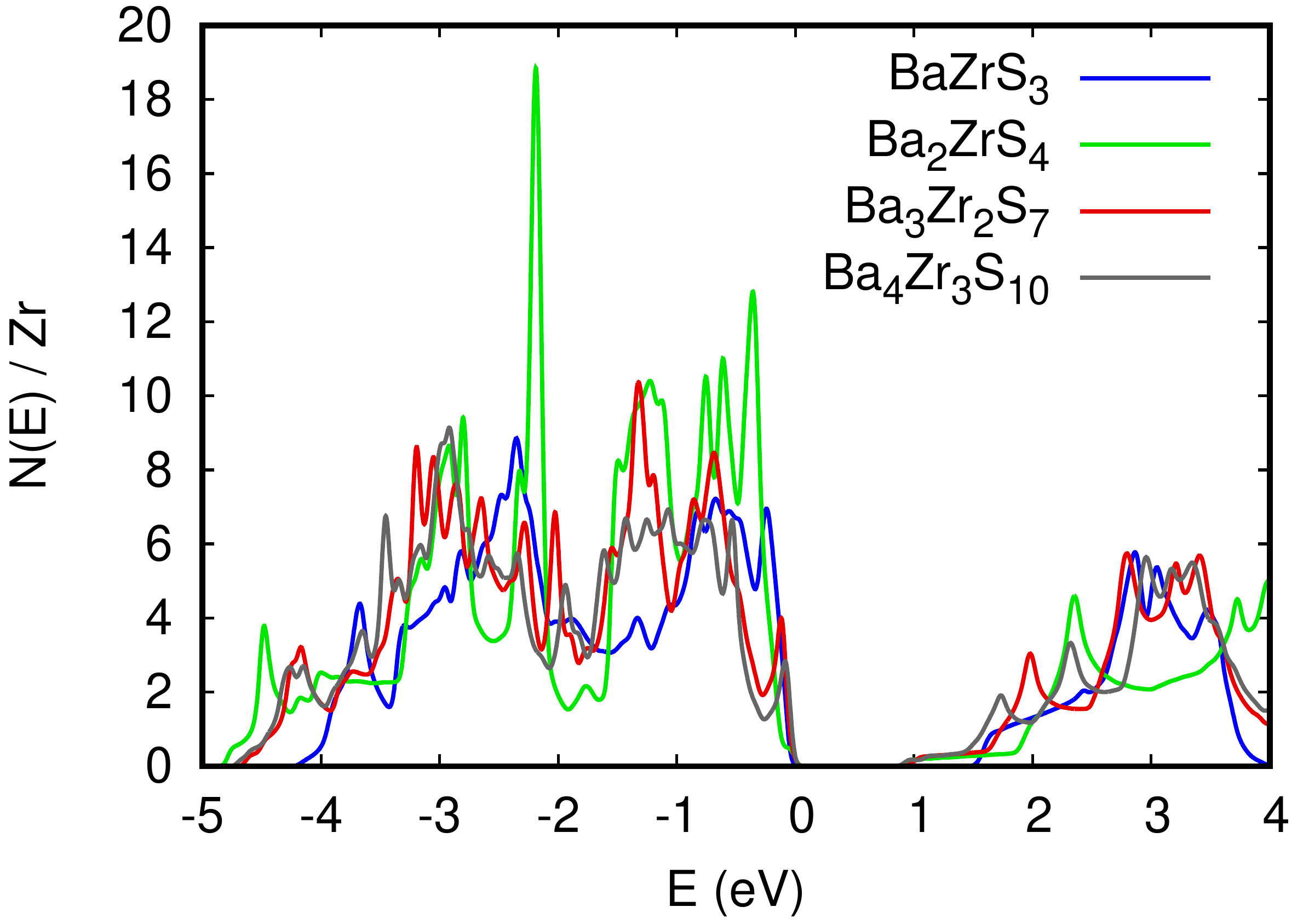}
\caption{Calculated electronic densities of states plotted on a per Zr basis
for the four compounds.}
\label{alldos}
\end{figure}

\begin{figure}
\includegraphics[width=0.95\columnwidth,angle=0]{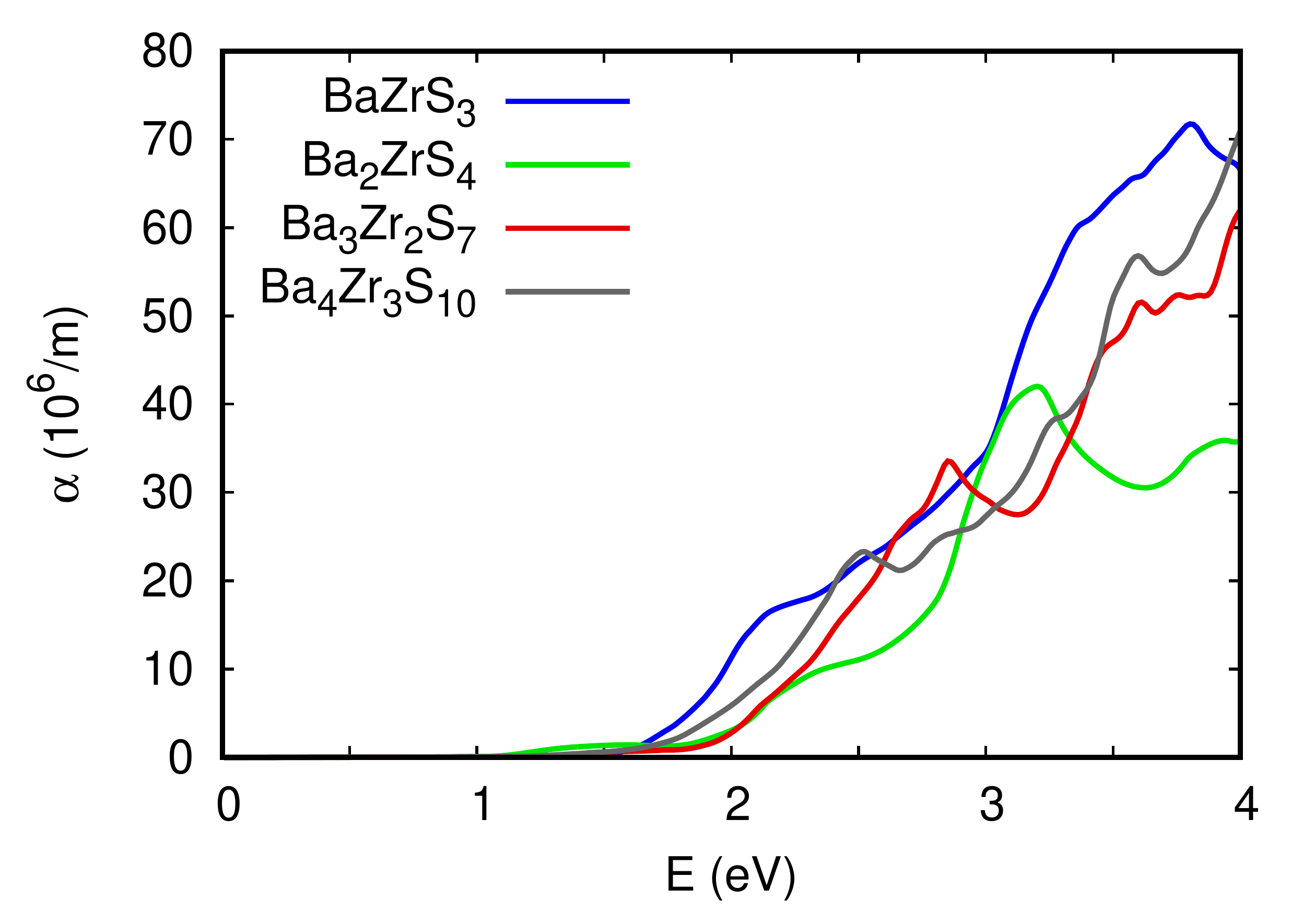}
\caption{Calculated direction averaged absorption spectra for the four 
compounds. A Lorentzian broadening of 0.025 eV was applied.}
\label{allabs}
\end{figure}

We begin with our electronic structure results.
The electronic densities of states (DOS) of the different compounds are
compared in Fig. \ref{alldos},
and the calculated direction averaged
optical absorption spectra are compared in
Fig. \ref{allabs}. The band gaps and onsets of optical
absorption are listed in Table \ref{bandgap}.
Rock salt structure BaS has an indirect band gap of 
3.32 eV based on the same method.
Thus we find BaS to be a wide band gap material, consistent with
other reports. \cite{zagorac}
The implication of the wide gap of BaS is that the rock salt parts of the
unit cells of the RP phases can be expected to serve as blocking layers
from an electronic point of view, so that the RP phases may be expected
to show 2D characteristics in their electronic structure.

\begin{table}[tb]
\caption{Crystal structures, including relaxed atomic positions, as
used in the calculations.}
\begin{tabular}{l c c c}
\hline
\hline
BaZrS$_3$ &  $Pnma$ \\
 &  $a$=7.0599 \AA,&  $b$=9.9813 \AA, &  $c$=7.0251 \AA  \\
\hline
 & $x$ & $y$ & $z$ \\
\hline
Ba &  0.4495 & 1/4 & 0.0098 \\
Zr &  0 & 0 & 0 \\
S1 &  0.5040 & 1/4 & 0.5673 \\
S2 &  0.2084 & 0.0348 & 0.2913 \\
\hline
\hline
Ba$_2$ZrS$_4$ &  $I4/mmm$ \\
 &  $a$=4.7852 \AA, &  $c$=15.9641 \AA \\
\hline
 & $x$ & $y$ & $z$ \\
\hline
Ba &   0 & 0 & 0.3560 \\
Zr &   0 & 0 & 0 \\
S1 &   0 & 1/2 & 0 \\
S2 &   0 & 0 & 0.1640 \\
\hline
\hline
Ba$_3$Zr$_2$S$_7$ &  $I4/mmm$ \\
 &  $a$=4.9983 \AA, & $c$=25.502 \AA \\
\hline
 & $x$ & $y$ & $z$ \\
\hline
Ba1 & 0   & 0   & 1/2 \\
Ba2 & 0   & 0   & 0.3193 \\
Zr  & 0   & 0   & 0.1007 \\
S1  & 0   & 1/2 & 0.0960 \\
S2  & 0   & 0   & 0.1993 \\
S3  & 0   & 0   & 0   \\
\hline
\hline
Ba$_4$Zr$_3$S$_{10}$ & $Fmmm$ \\
 &  $a$=7.0314 \AA, & $b$=7.0552 \AA, & $c$=35.544 \AA \\
\hline
 & $x$ & $y$ & $z$ \\
\hline
Ba1 & 0   &  0   &  0.4322 \\
Ba2 & 0   &  0   &  0.3001 \\
Zr1 & 0   &  0   &  0   \\
Zr2 & 0   &  0   &  0.1436 \\
S1  & 0   &  0   &  0.0708 \\
S2  & 0   &  0   &  0.2140 \\
S3  & 1/4  &  1/4  &  0   \\
S4  & 1/4  &  1/4  &  0.1395 \\
\hline
\hline
\end{tabular}
\label{struct-tab}
\end{table}

The DOS shows that the RP phases all have smaller band gaps than the
3D perovskite.
This is an important consideration for solar absorber applications, discussed
further below.
The onset of the DOS at the top of the valence bands
is similar for all the compounds in that the DOS rises sharply as the
energy is lowered into the valence bands. The DOS shapes for the conduction
bands of the RP phases show some
qualitative characteristics expected for a 2D material,
in particular, step like features as a function of energy.
It is also notable that the RP phases all show a low 
density of states tail extending from the
conduction band edge to an onset of much more steeply increasing
DOS at higher energy.

\begin{figure}
\includegraphics[width=0.95\columnwidth,angle=0]{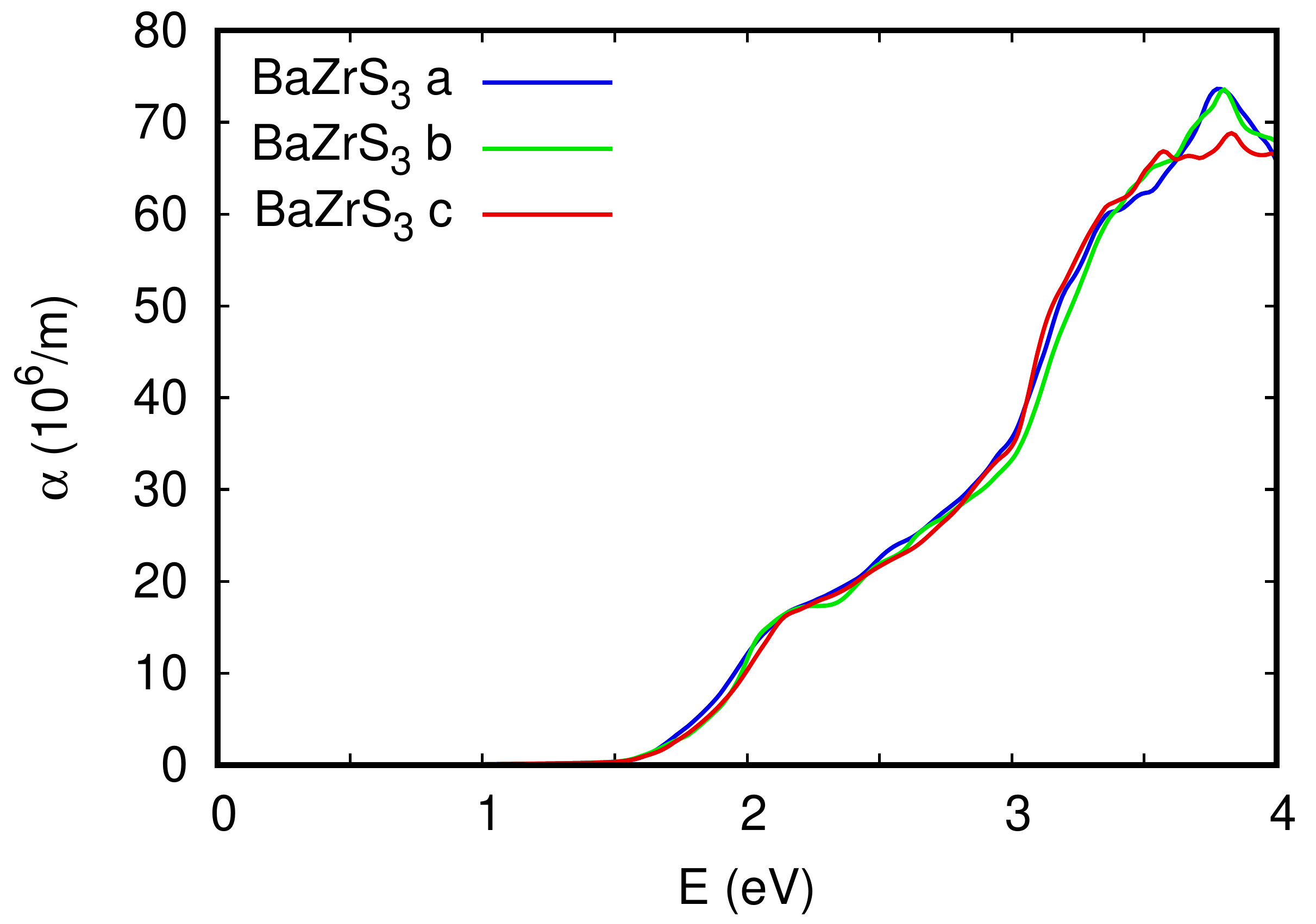}
\caption{Calculated direction dependent absorption spectra
of BaZrS$_3$. The labels 'a', 'b' and 'c' denote light polarization
along the $a$, $b$ and $c$ axes, respectively.
A Lorentzian broadening of 0.025 eV was applied.}
\label{abs113}
\end{figure}

BaZrS$_3$ has been discussed as a potential solar photovoltaic material.
\cite{sun}
BaZrS$_3$ has a 1.50 eV direct band gap at $\Gamma$.
The other compounds are indirect band gap materials, whose
indirect and optical gaps are given in Table \ref{bandgap}.
The present value for the band gap of BaZrS$_3$ is slightly lower
than our previously reported value of 1.55 eV, \cite{niu}
due to the inclusion of spin-orbit in the present calculation.
Our value is also smaller than reported hybrid functional values of
$\sim$1.7 eV -- 1.8 eV, without spin orbit. \cite{sun,meng,polfus}
Experimental band gap values are limited to powder photoluminescence
and diffuse reflectance measurements with Kubelka-Munk modeling.
The values obtained in this way are 1.85 eV, \cite{meng} 
1.7 eV, \cite{perera} and 1.83 eV. \cite{niu}
Our calculated direction dependent absorption spectra are given in
Fig. \ref{abs113}. It will be of interest to compare with
single crystal optical spectra, if suitable samples become
available.

\begin{figure}
\includegraphics[width=0.85\columnwidth,angle=0]{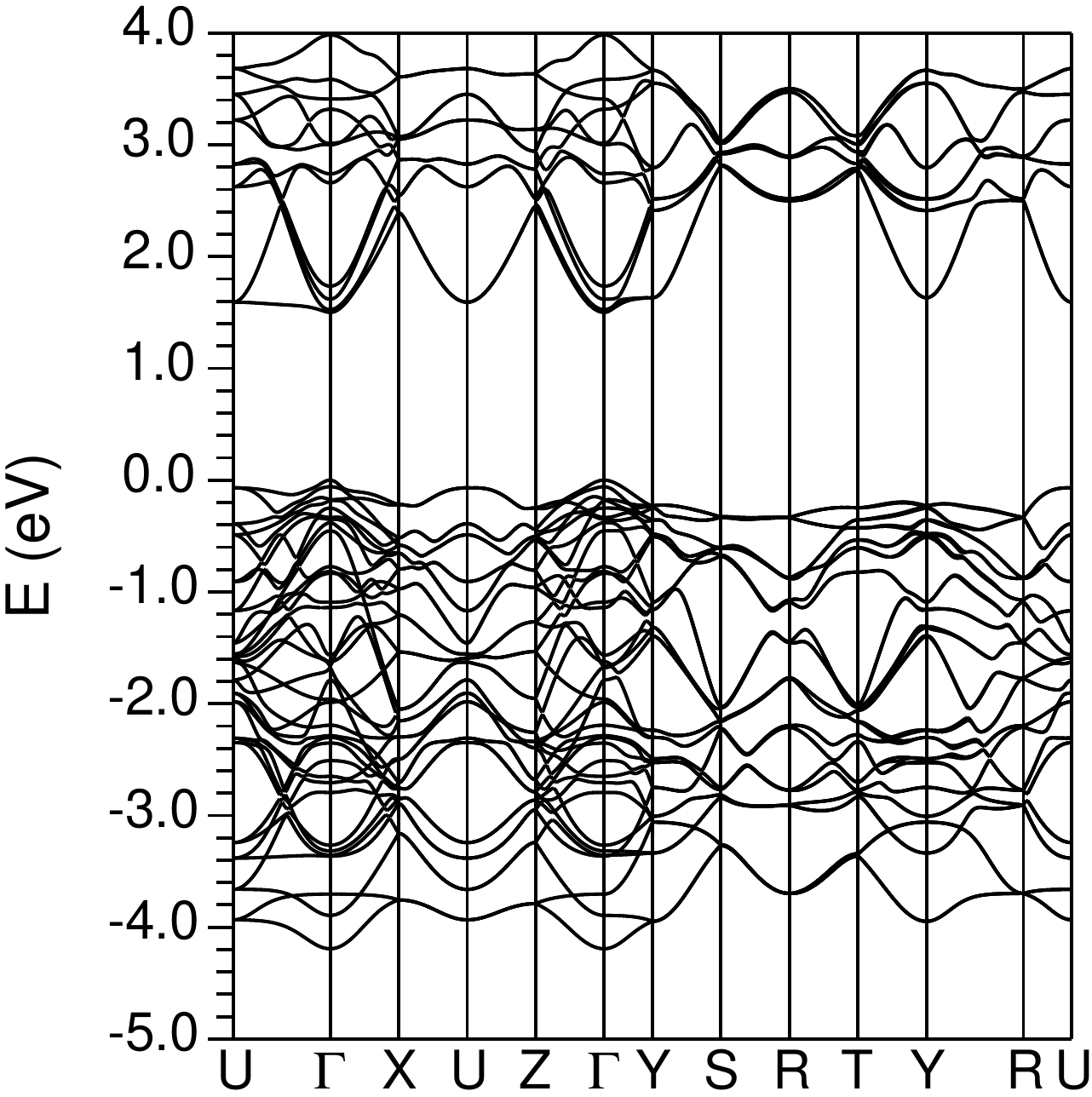}
\caption{Band structure of BaZrS$_{3}$ including
spin orbit. Symmetry labels follow the notation of Lax.\cite{lax-book}}
\label{band-113}
\end{figure}

Although BaZrS$_3$ is orthorhombic, the
absorption is rather isotropic over the whole energy range, which
should simplify the analysis of experimental data.
The absorption onset is at the direct band gap of 1.5 eV.
However, as may be seen there is a shoulder in the absorption spectrum
at $\sim$2.2 eV, and at energies below this shoulder the spectrum
is concave upwards, especially below 2 eV. 
Ordinary parabolic band direct gap semiconductors have absorption,
$\alpha$=$A(E-E_g)^{1/2}$, where $A$ a coefficient.
This form is concave downwards.
The band structure is shown in Fig. \ref{band-113}.
The unusual shape of the absorption edge
is a consequence of the non-parabolicity of the bands.
This is
due to the near degeneracy of the bands at the $\Gamma$-point valence
and conduction band edges, as well as the related characteristic
flat lowest conduction band (seen along $\Gamma$-U) which arises from the
shape of the $t_{2g}$ orbitals in relation to the ligand coordination
in a perovskite structure, as has been discussed elsewhere in relation
to transport and optical properties of other materials.
\cite{sun-sto}
In any case, this unusual shape of the optical absorption
may then explain fits of optical spectra
extrapolating to gap values near 1.8 eV, with the lower energy
part of the spectrum perhaps mis-characterized as an Urbach tail.
It will be of interest to perform quantitative analysis using
single crystal absorption or reflection data should suitable
samples become available.

\begin{table}
\caption{Indirect and optical band gaps of the different compounds as obtained
using the mBJ potential including spin-orbit.}
\begin{tabular}{lcc}
\hline
Compound~~~~   & $E_g$(ind.) (eV) ~~~ & ~~~ $E_g$(opt.) (eV) \\
\hline
BaZrS$_3$            & 1.50 & 1.50 \\
Ba$_2$ZrS$_4$        & 0.92 & 1.05 \\
Ba$_3$Zr$_2$S$_7$    & 0.95 & 1.26 \\
Ba$_4$Zr$_3$S$_{10}$ & 0.88 & 1.20 \\
\hline
\end{tabular}
\label{bandgap}
\end{table}

From the point of view of materials for photovoltaic application,
the Shockley-Queisser maximum single junction efficiency curve \cite{shockley}
is essentially flat at the maximum value from $\sim$1 eV -- 1.5 eV.
Thus these materials all have suitable band gaps for solar applications.
However, except for BaZrS$_3$, the band gaps are indirect, which is generally
unfavorable for this application.
The differences between the direct and indirect band gap range
from 0.13 eV in Ba$_2$ZrS$_4$ to $\sim$0.3 eV for the other
two RP phases. The relatively smaller difference between the indirect
and optical band gaps for Ba$_2$ZrS$_4$ and its suitable gap value
mean that this materials may be the most likely of the RP phases to be a
useful solar absorber, with the Ba$_3$Zr$_2$S$_7$ phase next.

\begin{figure}
\includegraphics[width=0.85\columnwidth,angle=0]{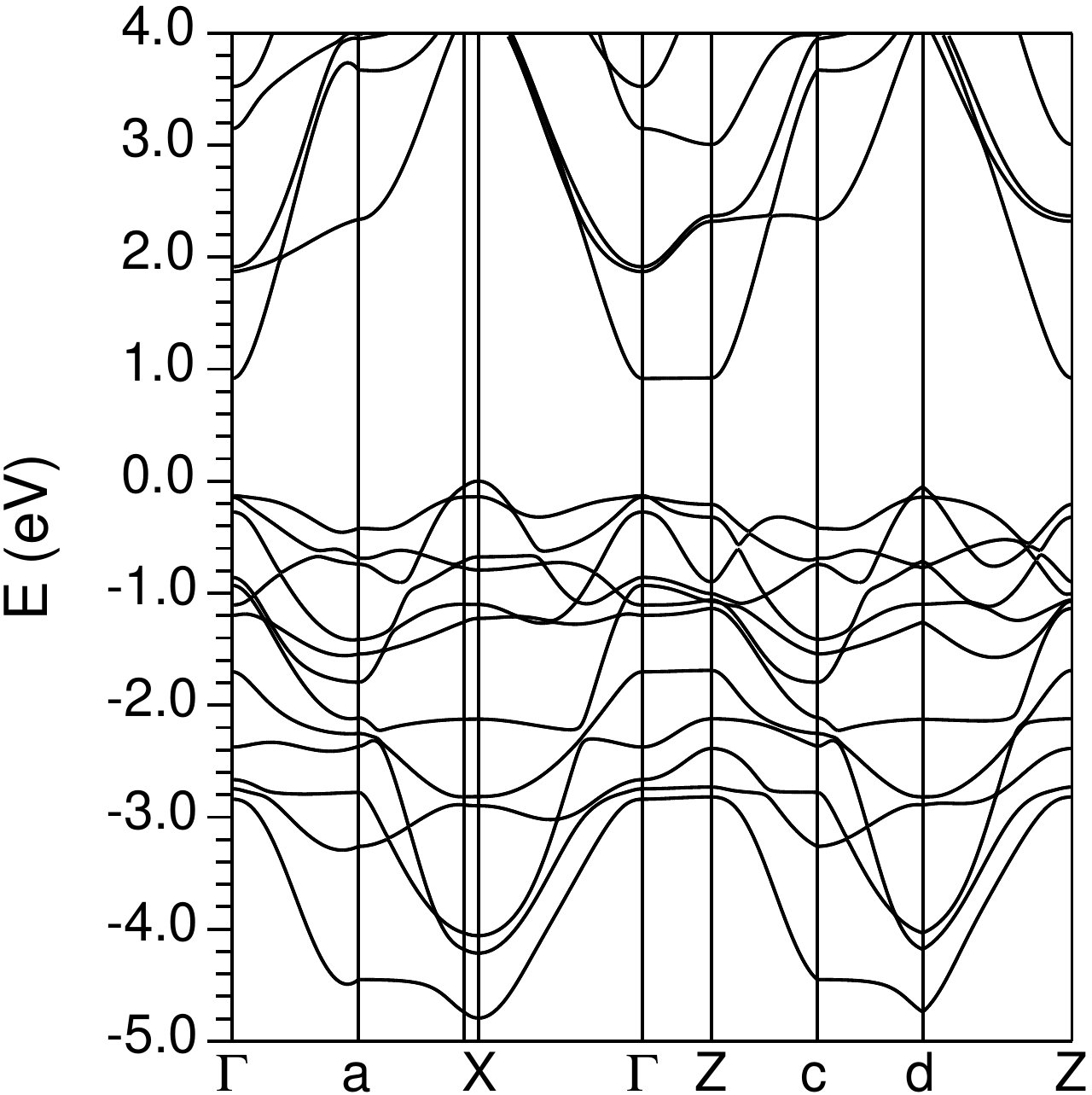}
\caption{Band structure of Ba$_2$ZrS$_{4}$ including
spin orbit.
Symmetry points are labeled using
capitals according to the notation of Lax. \cite{lax-book}.
Non-symmetry points are indicated with lower case letters.
The Brillouin zone and band structure path are shown in Fig. \ref{band-path}.
Note that the point 'b' is immediately before X in the path and is
not marked on the axis due to space.}
\label{band-214}
\end{figure}

\begin{figure}
\includegraphics[width=\columnwidth,angle=0]{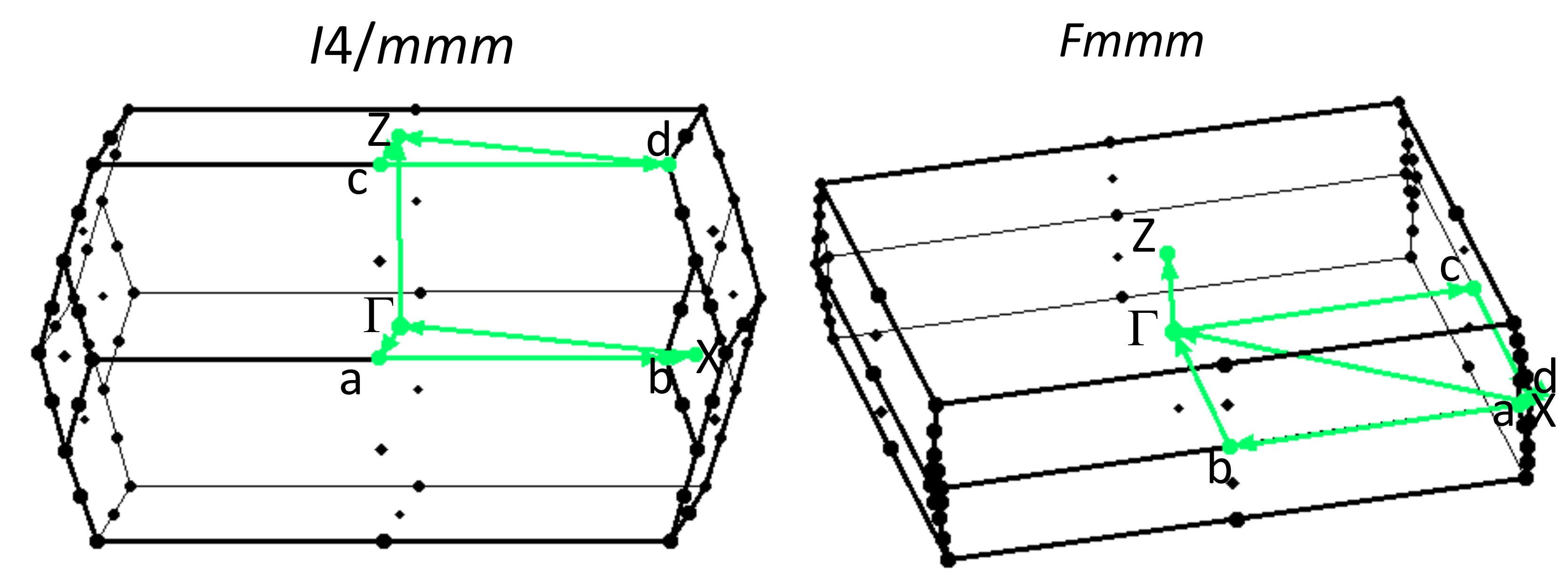}
\caption{Brillouin zone and path through the zone
for the $I4/mmm$ and $Fmmm$ structures.}
\label{band-path}
\end{figure}

The band structure of Ba$_2$ZrS$_4$ is shown in Fig. \ref{band-214}.
The path through the zone is shown in Fig. \ref{band-path},
and is chosen to have lines either in the basal plane (constant $k_z$)
or perpendicular to it. This is in order to more clearly show the relative
dispersions in-plane and out-of-plane.
The lowest conduction band is very flat along $\Gamma$-Z, meaning that
it is a very 2D band. In contrast, it is very dispersive in-plane.
This band is single degenerate, neglecting spin, and comes
from the xy orbital of the Zr $t_{2g}$ manifold.
It should be noted that this orbital points in the layer plane of the
K$_2$NiF$_4$ structure, consistent with the very weak $k_z$ dispersion
of this band.
The next conduction band at $\Gamma$ is two-fold degenerate, corresponding
to the xy/xz $t_{2g}$ orbitals. This band is seen to be more dispersive
along $k_z$ ($\Gamma$-Z), reflecting the fact that the lobes of
the underlying orbitals do not point in-plane.
These bands are also less
dispersive than the lowest band. This lower dispersion corresponds to the
smaller band width. This smaller band width explains why the
band minimum for these
two bands lies higher in energy than the bottom of the xy-band. This
smaller width is simply understood if one considers the xy orbital, which
points in-plane and so has hopping in both the x and y directions,
in relation to e.g. the xz orbital, which does not have as much hopping in the
y direction.

The DOS of an ideal
2D parabolic band is in the shape of a step function, with height inversely
proportional to the effective mass.
This is the explanation of the low value DOS tail at the conduction
band edge.
The top valence band is more dispersive along $\Gamma$-Z, as are
bands further from the band edges.

\begin{figure}
\includegraphics[width=0.85\columnwidth,angle=0]{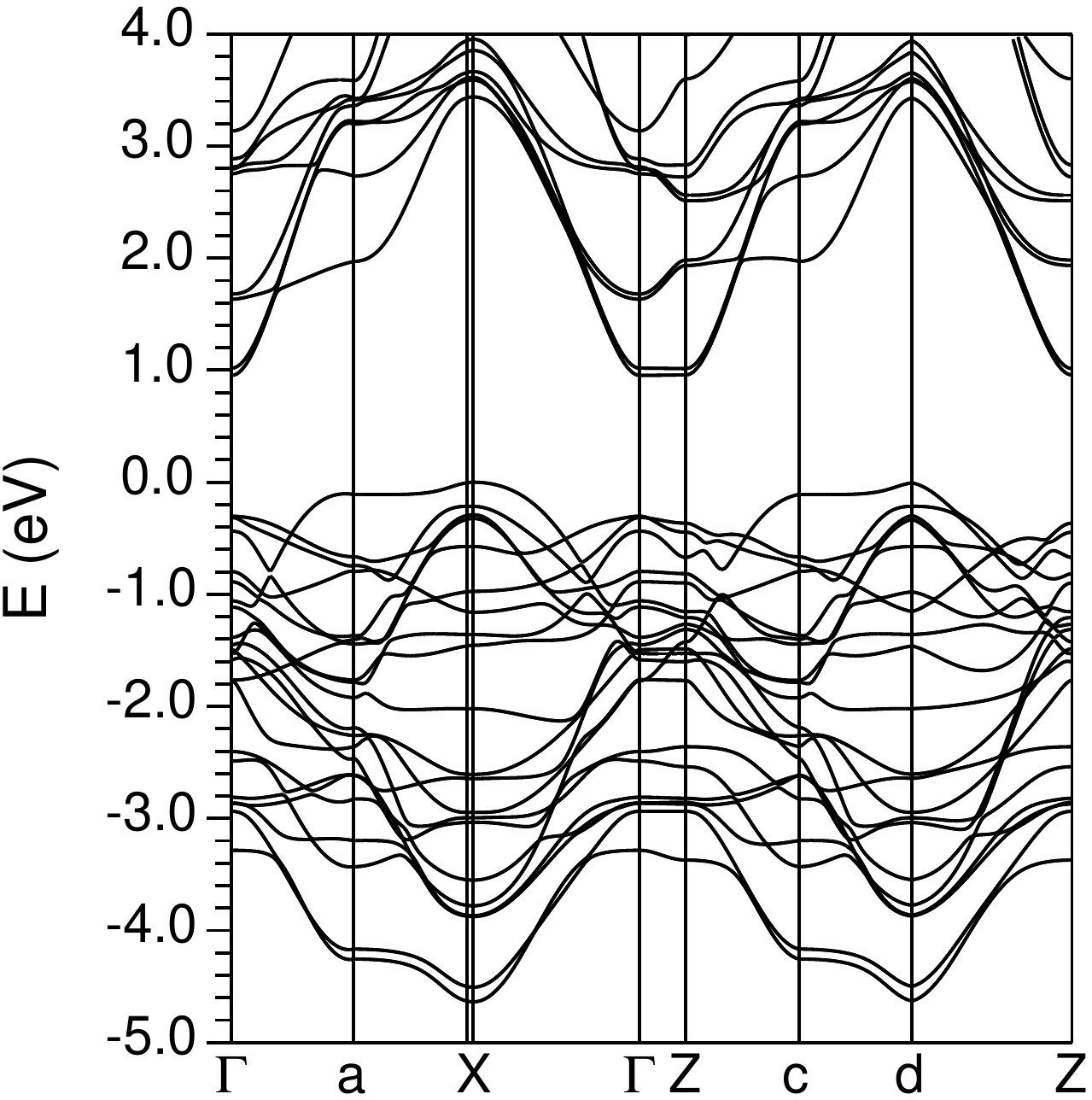}
\caption{Band structure of Ba$_3$Zr$_2$S$_{7}$ including
spin orbit. The k-point path is the same as in Fig. \ref{band-214}.}
\label{band-327}
\end{figure}

\begin{figure}
\includegraphics[width=0.85\columnwidth,angle=0]{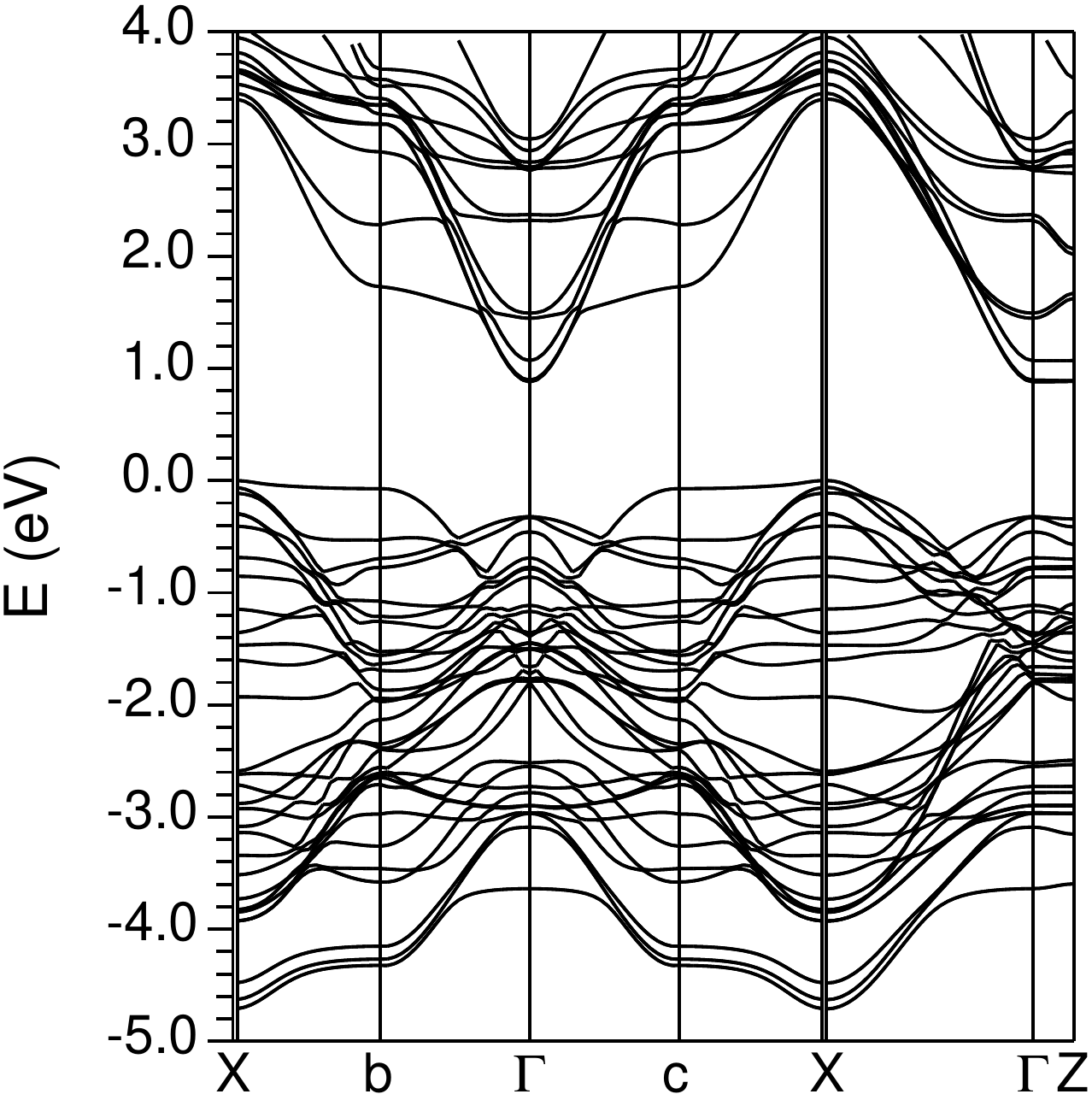}
\caption{Band structure of Ba$_4$Zr$_3$S$_{10}$ including
spin orbit.
The Brillouin zone and band structure path are shown in Fig. \ref{band-path}.
Symmetry points are labeled using
capitals according to the notation of Lax. \cite{lax-book}.
Non-symmetry points are indicated with lower case letters.
Note that the point 'a' is immediately after the first X 
and 'd' is immediately before the second X in the path and is
not marked on the axis due to space.}
\label{band-4310}
\end{figure}

The band structure of the $n$=2 compound, Ba$_3$Zr$_2$S$_7$
(Fig. \ref{band-327}),
which also occurs in an ideal tetragonal structure without octahedral
rotation, can be understood in the same way.
In particular, there are now two slightly split bands at the
bottom of the conduction band at $\Gamma$. These are associated with
the xy orbitals of the two layers in the structure, and are weakly
split because of the in-plane nature of the xy orbital.
The four xz/yz bands interact more strongly and are accordingly
much more strongly split at $\Gamma$, in addition to showing stronger
$k_z$ dispersion. In any case, the xy bands form the conduction band
minimum similar to the $n$=1 compound, and again there is a low flat
tail on the DOS at the bottom of the conduction band.
The $n$=3 compound, Ba$_4$Zr$_3$S$_{10}$ has a lower symmetry distorted
orthorhombic crystal crystal structure. This leads to a more complex
band structure, as shown in Fig. \ref{band-4310}.
However, the qualitative
features of an indirect band gap, with a valence band maximum at the
zone corner, a conduction band minimum at the zone center,
and a conduction band minimum derived from bands with low dispersion
along the $k_z$ direction remain.

\begin{figure}
\includegraphics[width=0.95\columnwidth,angle=0]{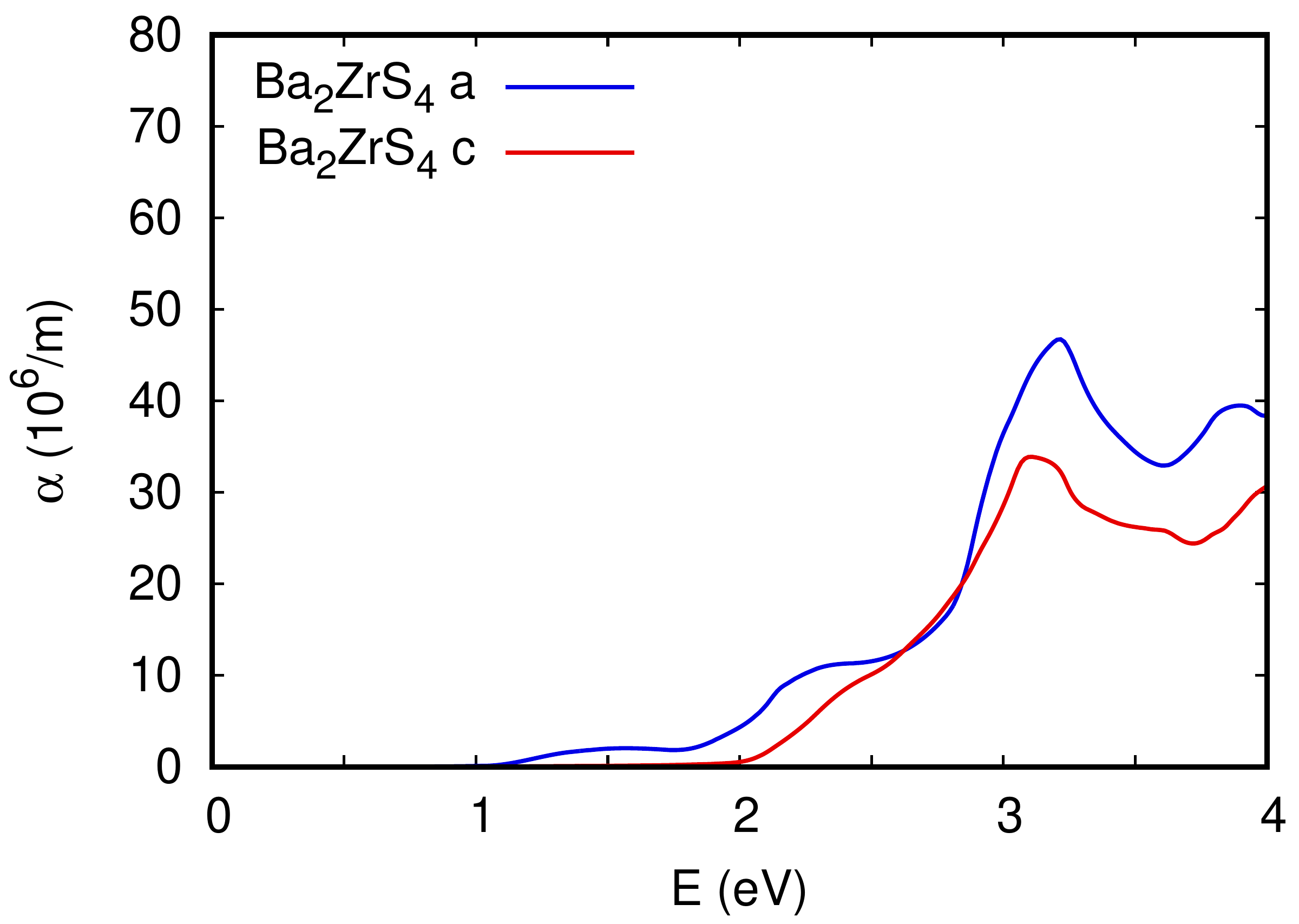}
\caption{Optical absorption spectrum of Ba$_2$ZrS$_4$ for
polarization in-plane, denoted 'a', and along the $c$-axis,
denoted 'c'.}
\label{abs-214}
\end{figure}

\begin{figure}
\includegraphics[width=0.95\columnwidth,angle=0]{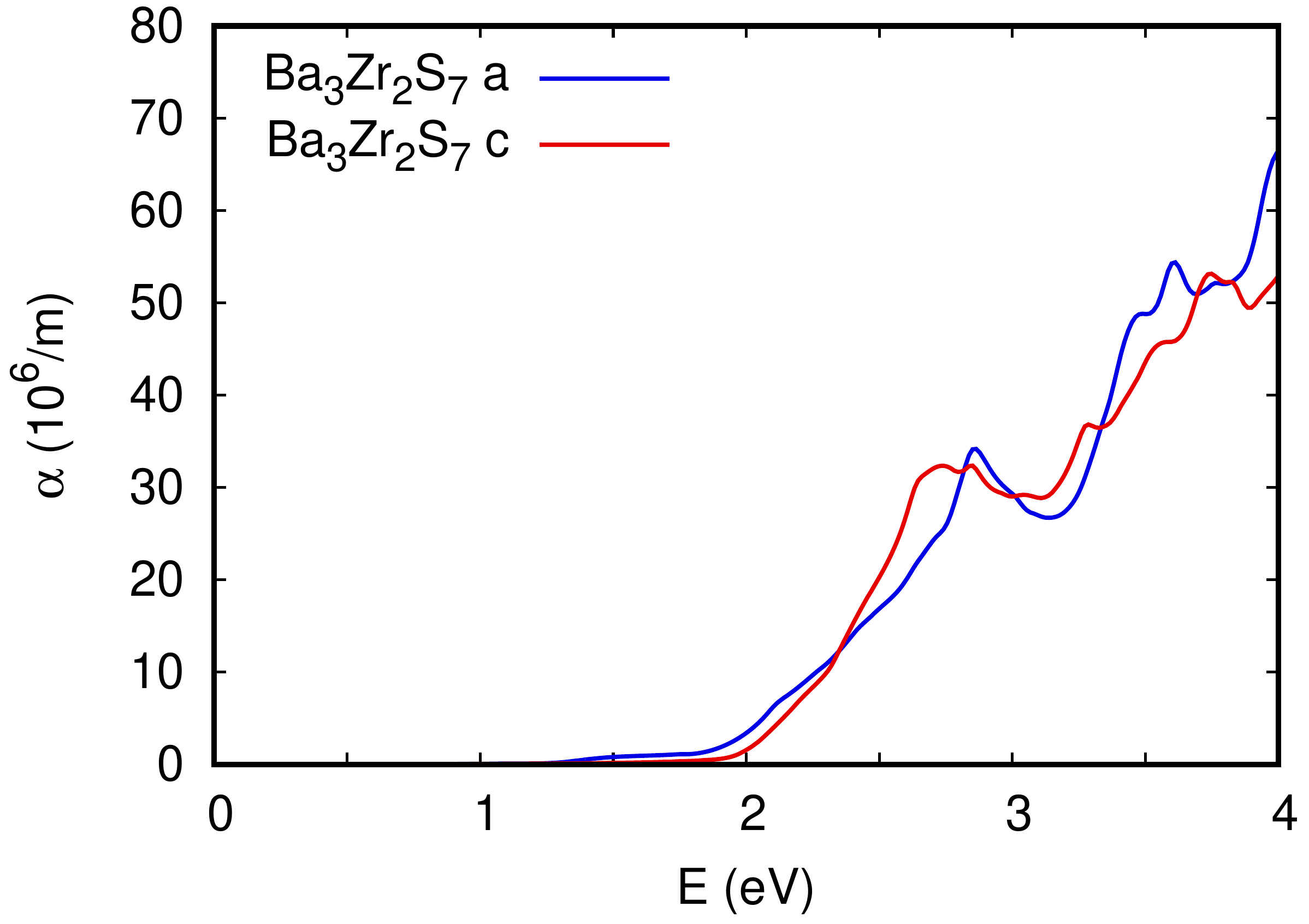}
\caption{Optical absorption spectrum of Ba$_3$Zr$_2$S$_7$ for
polarization in-plane, denoted 'a', and along the $c$-axis,
denoted 'c'.}
\label{abs-327}
\end{figure}

\begin{figure}
\includegraphics[width=0.95\columnwidth,angle=0]{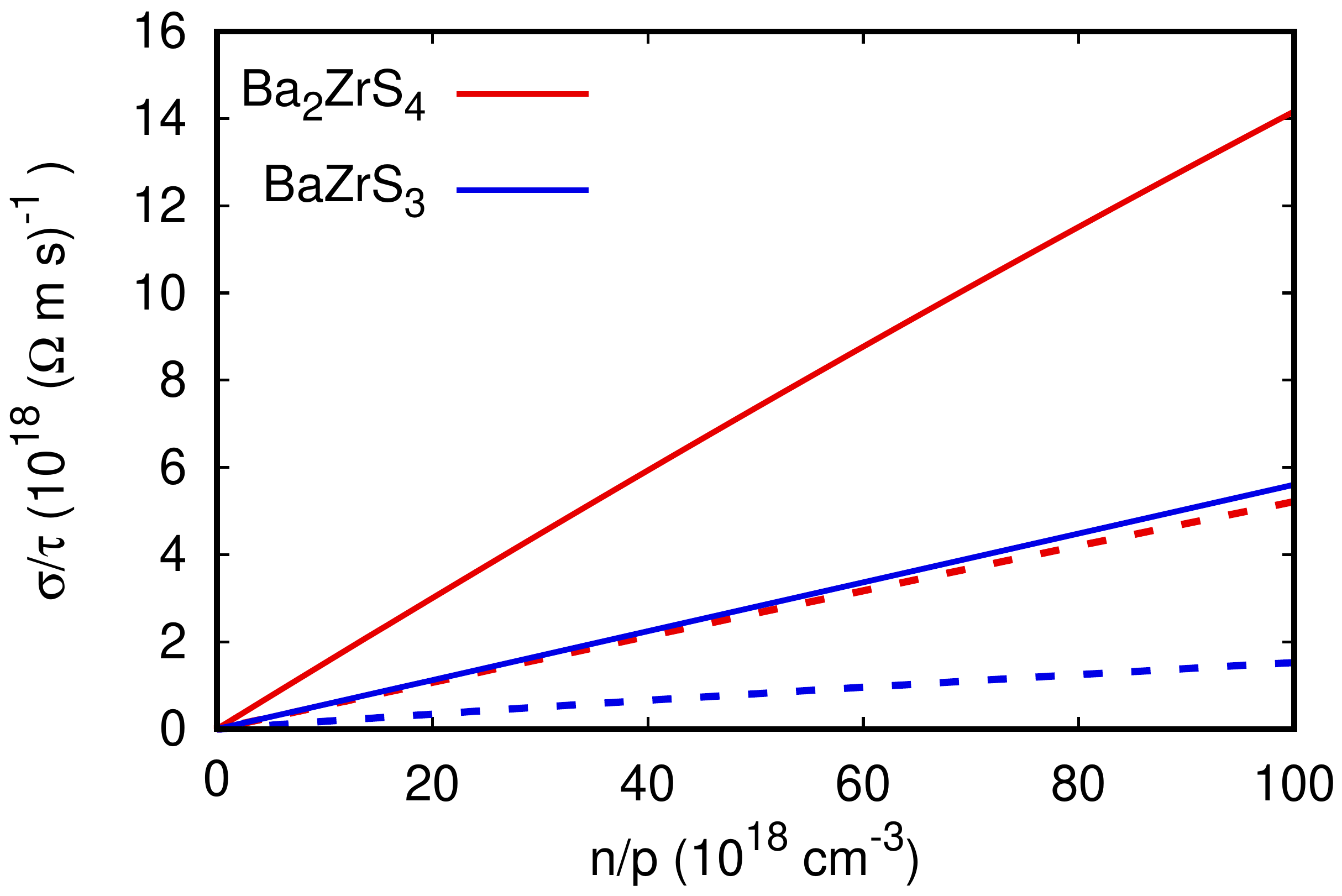}
\caption{Transport function $\sigma/\tau$ for in plane transport
in Ba$_2$ZrS$_4$, in-plane (red), and BaZrS$_3$, direction averaged (blue),
as functions of carrier concentration. n-type is shown using solid lines
and p-type with dashed lines.
}
\label{sigma}
\end{figure}

Importantly, in both the $n$=1 and $n$=2 compounds the valence band
maximum is at the zone corner X point, while the conduction band
minimum is at $\Gamma$, as discussed above. The optical gap is at the
$\Gamma$ point. This makes the low tail in the DOS also of relevance for
the optical absorption spectra,
shown in Figs. \ref{abs-214} and \ref{abs-327} for Ba$_2$ZrS$_4$ and
Ba$_3$Zr$_2$S$_7$, respectively.
As may be seen, these features of the band
structure are reflected in what might be termed a polarization
dependent optical gap. What is meant by this is that the xy nature of the
conduction band minimum allows optical transitions at $\Gamma$
for in-plane light polarization, but not for c-axis polarized light.
For example, in Ba$_2$ZrS$_4$ we find that the absorption spectrum
for c-axis polarized light goes smoothly to zero with an apparent
optical band bap of $\sim$2 eV. On the other hand, for in-plane polarization
the absorption extends to the optical gap of 1.05 eV. It is also
important to note that the absorption in this foot is relatively low,
and therefore, especially in direction averages, (see Fig. \ref{allabs})
could easily be mistaken for an extrinsic tail, leading to the assignment
of a higher than actual band gap.

A second consequence of the dimensional reduction
is that the optical spectrum is distinctly anisotropic, not only near
the band edge as discussed above, but also including significantly 
anisotropic absorption to above 4 eV in Ba$_2$ZrS$_4$.
Ba$_3$Zr$_2$S$_7$ is also anisotropic near the band edge, but
is less anisotropic at higher energy, presumably due to the
more bulk-like bilayer perovskite block in this compound.

Finally, it of interest to consider the effect of dimensional
reduction on the band structures in relation
to carrier transport. In general, this requires knowledge of scattering
mechanisms for the different compounds and samples along with the
temperature dependencies.
However, a comparison can be made based on the band structure.
Specifically, within the relaxation time approximation for the linearized
Boltzmann transport theory, the conductivity, $\sigma$ can be expressed
as $\tau \sigma/\tau$, where $\tau$ is a relaxation time and
$\sigma/\tau$ depends only on the band structure and temperature.
We used the BoltzTraP code \cite{boltztrap} to calculate this function
at 300 K for in-plane transport in Ba$_2$ZrS$_4$ as compared to that
in the 3D perovskite, BaZrS$_3$, which is nearly isotropic and
for which a direction
average is accordingly shown (Fig. \ref{sigma}).
Both compounds have higher $\sigma/\tau$ for electrons than
for holes, as might be anticipated from the more dispersive conduction
bands, relative to the valence bands.
Significantly,
the dimensional reduction greatly increases $\sigma/\tau$ for
Ba$_2$ZrS$_4$ relative to BaZrS$_3$ even though the dimensional
reduction was accomplished by adding a layer of insulating BaS to the
unit cell. Furthermore this enhancement of in-plane transport is clear both
for electrons and holes.

\section{Summary and Conclusions}

First principles calculations for the Ba-Zr-S, RP series show that
there is considerable tunability of the electronic and optical
properties of BaZrS$_3$ via dimensional reduction.
This includes a transition from
a direct gap for the perovskite to indirect gaps for the RP phases
and reduction of the band gap by up to 0.6 eV.
While the perovskite shows a nearly isotropic optical
absorption, the RP phases are distinctly anisotropic and different
from each other, especially at energies near the absorption onset.
In the context of photovoltaics, among the RP phases, Ba$_2$ZrS$_4$
is favorable in terms of band gap, since the difference
between the indirect and direct band gaps is relatively small, and
the loss of voltage due to the indirect gap might be partly compensated
for by an increase in minority carrier lifetime due to suppression
of recombination related to the indirect gap. Furthermore, this compound
is superior to the 3D perovskite, BaZrS$_3$ from the point of view of
in-plane carrier transport. Therefore it is possible that suitably
oriented
Ba$_2$ZrS$_4$ may be of interest in photovoltaics, in
addition to the 3D counterpart BaZrS$_3$.
Investigation of the Ba$_3$Zr$_2$S$_7$ would also be of interest.
In any case, the substantial electronic tunability by dimensional
reduction in these phases may be of important for other electronic
applications.

\acknowledgments

We acknowledge useful discussion with Jayakanth Ravichandran.
This work was supported by the Department of Energy, through the
S3TEC Energy Frontier Research Center, Award DE-SC0001299.
This is work is dedicated to Professor E.K.U. Gross.

\section*{Author Contribution Statement}

Calculations were done by DJS and YL.
The manuscript was written by DJS.

\bibliography{BaZrS3}

\end{document}